\documentclass[a4paper,11pt]{iopart}
\pdfoutput=1
\usepackage{graphicx}
\usepackage{amsfonts,amsthm,bm}
\usepackage{epstopdf}

\newcommand{\RGf}{\ensuremath{\mathcal{R}}}
\newcommand{\ave}[1]{\langle #1 \rangle}

\begin{document}

\title[Self-avoiding walks and biased differential approximants ]{Square lattice self-avoiding walks and biased differential approximants}
\author{Iwan Jensen }
\address{School of Mathematics and Statistics,
The University of Melbourne, VIC 3010, Australia}

\ead{ij@unimelb.edu.au} 
\date{\today}                                           

\begin{abstract}
The  model of self-avoiding  lattice walks and the asymptotic analysis of power-series have 
been two of the major research themes of Tony Guttmann. In this paper we bring the two
together and perform a new analysis of the generating functions for the number of square lattice
self-avoiding walks and some of their metric properties  such as the mean-square end-to-end distance.
The critical point $x_c$ for self-avoiding walks is known to a high degree of accuracy and we utilise 
this knowledge to undertake a new numerical analysis of the series using biased differential approximants. 
The new method is major advance in asymptotic power-series analysis in that it allows us to bias 
differential approximants to have a singularity of order $q$ at $x_c$. When biasing at $x_c$
with $q\geq 2$ the analysis yields a very accurate estimate for the critical exponent $\gamma=1.3437500(3)$ 
thus confirming the conjectured exact value $\gamma=43/32$ to 8 significant digits and removing 
a long-standing minor discrepancy between exact and numerical results. The analysis of the 
mean-square end-to-end distance yields $\nu=0.7500002(4)$ thus confirming
the exact value $\nu=3/4$ to 7 significant digits.
\end{abstract}

\noindent {\bf PACS}: 05.50.+q, 05.10.-a, 02.60-x

\noindent 
{\bf MSC}: 05A15,  30B10, 65D15, 82B20, 82B27, 82B41

\noindent
{\bf Keywords:} Self-avoiding walks, critical exponents, power-series expansions, asymptotic series analysis

\section{Introduction}

In Tony Guttmann's long and distinguished career the model of self-avoiding walks
and how to analyse its behaviour has been a mainstay. 
Since the early years \cite{Sykes72}
to the present day \cite{Clisby2016} Tony has published more than 130 papers on
self-avoiding walks, polygons and closely related models. At the same time Tony has made many
important and seminal contributions to the development of new methods for asymptotic analysis of power-series,
chief amongst these the method of differential approximants \cite{Guttmann72,GuttmannDA}.

A {\em $n$-step self-avoiding walk} (SAW) ${\bf \omega}$ on a regular lattice is 
a sequence of {\em distinct} vertices $\omega_0, \omega_1,\ldots , \omega_n$ 
such that each vertex is a nearest neighbour of it predecessor. SAW are
considered distinct up to translations of the starting point $\omega_0$.
If  $\omega_0$ and  $\omega_n$ are nearest-neighbours we can form
a closed polygon $n$-step self-avoiding polygon (SAP) by adding
an edge between the two end-points.
The fundamental problem is the calculation of the number 
of SAW, $c_n$, with $n$ steps. As most interesting combinatorial problems, 
SAW have exponential growth. 
The generating function for SAW (and SAP) is believed to have algebraic singularities,  
though in most cases this has not been proved. That is to say, the generating function 
is believed to behave as
\begin{equation}\label{eq:Gf}
F(x) \;=\; \sum_{n=0}^{\infty} c_xx^n \;\sim\; A(1 - x/x_c)^{-\gamma} \,\, {\rm as} \,\, x \to x_c^-.
\end{equation}
Here $A$ is referred to as the   {\it critical amplitude,} 
$x_c$ as the {\it critical point,} 
and $\gamma$ as the {\it critical exponent.}  The value of $\gamma = 43/32$ is known exactly \cite{Nienhuis82,Nienhuis84},
though it remains to be proved rigorously. Besides the physical singularity  there is another 
singularity at $x=x_-=-x_c$ \cite{Guttmann78,Barber78} which has a
critical exponent consistent with the exact value $1/2$.

Before one can say much about the behaviour of self-avoiding walks one must have
something to work with so Tony has pushed hard for major strides to be made in the development of
new and improved algorithms \cite{Conway93a} for counting self-avoiding walks. 
And given that these counting problems are inherently of exponential time complexity Tony and his co-workers have
always been  eager to make use of the latest progress in computing technology
including being early adapters of massively parallel computations \cite{Conway96}. The data for self-avoiding walks
has been used to obtain very accurate numerical tests of analytic predictions originating 
from scaling theory, conformal field theory, stochastic Loewner evolution, etc.
\cite{Barber78,Caracciolo05,Guttmann13a}. Besides   the already mentioned method of differential approximants the intricacies of 
the asymptotic behaviour of various walk problem has spurred Tony's development of new methods 
for asymptotic analysis of power-series  \cite{Beaton10a,Beaton12b,Guttmann15}. The full self-avoiding walk model
is well know to be a very tough problem with no immediate prospect of an exact solution. In fact
Tony used SAW as a prime testing ground for ideas on how to determine whether or not
a given model is solvable \cite{Enting96,Guttmann00a}. Tony
has been an  avid advocate for the utility of studying simpler exactly solvable problem both as
an important and interesting pursuit in its own right and as a means of gaining further insight
into the original harder problem. This work has resulted in many
exact solutions to simpler (often directed) walk problems \cite{Guttmann88a,Brak90,Essam95,Guttmann98a} 
including one of the very few exact solutions for a 3D lattice model \cite{BousquetMelou97a,BousquetMelou97b}.
One particular result vividly illustrates why the study of simple solvable lattice models can
contribute in a major way to our understanding of more complicated problem. In a series
of papers Tony and co-workers demonstrated that the study of the simple model of staircase
polygons could lead one to conjecture the area-perimeter scaling function for self-avoiding polygons 
\cite{Richard01a,Richard01b,Richard08}. Finally, we mention that Tony has been involved
in many research projects which have used self-avoiding walk to model many aspects
of polymer physics and chemistry including steric stabilisation \cite{Guttmann78a}, modelling of
vesicles \cite{Fisher91}, polymer collapse and interacting walks \cite{Brak91,Brak92,Owczarek94,BennettWood98},
polymer adsorption and desorption from a surface \cite{Beaton12b,Beaton14a}, 
force-induced polymer unfolding \cite{Kumar07,Guttmann09} and desorption \cite{Guttmann14a}. 

So one can safely say  that self-avoiding walks have been very good to Tony and  that Tony has been 
exceptionally good for our understanding of the self-avoiding walk problem and its many and varied uses
in the modelling of physical, chemical and biological systems.

In this paper we bring together two of Tony's favourites and perform a new numerical analysis of 
the generating functions for the number of square lattice self-avoiding walks and some of their 
metric properties.  From the numerical analysis (using differential approximants) of the 
generating function for square lattice self-avoiding polygons we have obtained very accurate estimates 
for the critical point $x_c^2 =0.143680629269(2)$\footnote{The critical point for square lattice SAP is 
at $x_c^2$ because every polygon has even length.}  and critical exponent $\alpha=0.500000015(20)$ \cite{Clisby12}.
However, for  square lattice SAW there is an annoying (at least to the author) minor discrepancy
between the above prediction for the exact values of the critical exponents and the 
estimates from series analysis where our best estimate using standard differential approximants 
$\gamma=1.343745(3)$ is agonisingly close to the exact value $\gamma=43/32=1.34375$. Now 
no one would seriously take this minor discrepancy as an indication that the conjectured exact value isn't 
correct, rather it is a  `deficiency' in the numerical analysis. In this paper we show how this discrepancy 
can be eliminated by the use of {\em biased differential approximants}. With this new analysis we  can 
confirm the value of the critical exponent to at least 8 significant digits. We also carry out a 
biased analysis of the generating functions for the metric properties (end-to-end distance, 
monomer-to-end distance and radius of gyration) that confirms to 7 digits accuracy that the size exponent $\nu=3/4$.

\section{Biased differential approximants}

From the known exact solutions to various directed walk and polygon problems 
it is clear that the generating functions are often algebraic or given by the solution of simple linear 
ordinary differential equations \cite{Guttmann88a,Brak90,Essam95,Guttmann06a}. This observation 
(originally made in the context of the Ising model) forms the nucleus of the method of 
{\em differential approximants}. The basic idea is to approximate the function $F(x)$ by solutions
to differential equations with polynomial coefficients. The singular behaviour
of such ODEs is a well known classical mathematics problem 
(see e.g. \cite{Forsyth02,Ince56}) and the singular points and
exponents are easily calculated. Even if the function {\em globally} is not a solution
of a such a linear ODE (as is the case for SAW) one hopes that 
{\em locally} in the vicinity of the (physical) critical points the generating 
function can still be well approximated by a solution to a linear ODE.

A $K$th-order differential approximant (DA) to a function $F(x)$  is formed 
by representing the function by a $K$th-order ODE (possibly inhomogenous) 
with polynomial coefficients such that
\begin{equation} \label{eq:DA}
P(x)+\sum_{k=0}^K Q_{k}(x)\left(x\frac{{\rm d}}{{\rm d}x}\right)^k F(x)   = O(x^{N+1}).
\end{equation}
Here  $Q_k(x)$ and $P(x)$ are polynomials of degree $N_k$ and $L$, respectively, and
$$N=L+1+\sum_{k=0}^K (N_k+1)$$
is the number of  unknown coefficients. Determining these coefficients such that (\ref{eq:DA})  is
satisfied then amount to solving a system of linear equations.
 So we have that (one) of the formal power-series solutions 
to the ODE agrees with the  series expansion of the (generally unknown) 
function $F(x)$ up to the first $N$ coefficients. 

From the theory of ODEs, the singularities of $F(x)$ are approximated by the roots, 
$x_i \,\, (i=1, \ldots , N_K)$, of $Q_K(x),$ and the 
associated critical exponents $\lambda_i$ are estimated from the indicial equation. 
The physical critical  point $x_c$ is generally the singularity on the positive real axis closest to the origin.
When the root at $x_i$ has order $q$  we can find the associated exponents
by forming the indicial polynomial
\begin{equation}  \label{eq:indpol}
P_I(z)= \sum_{m=0}^q \, \frac{x_i^m}{m!}\cdot Q_{K-q+m}^{(m)}(x_i)\cdot  [z]_m,
\end{equation}
where $[z]_m = z(z-1)\cdots (z-m+1), \, [z]_0=1$. Then if $z_j,\; (j=1,\cdots,q)$ are the roots of $P_I(z)$ 
the associated critical exponents have values   $\lambda_{i,j}=K-q+z_j$. 
If there is only a single root at $x_i$  this is just
\begin{equation}  \label{eq:ana_indeq1}
\lambda_i=K-1-\frac{Q_{K-1}(x_i)}{x_iQ_K ' (x_i)}.
\end{equation}

By varying the degrees of the polynomials one can generate a large set of differential approximants to $F(x)$. 
Each approximant yields values for $x_c$ and $\gamma$ and by averaging over several (often hundreds)
of these one can calculate estimates for these critical parameters (see \cite[Ch 8]{PolygonBook} for more details).
We denote individual approximants by the notation $[N_K,N_{K-1},\ldots,N_0;L]$\footnote{Note that in the $L=0$
case  $P(x)=0$ rather than a non-zero constant.}. In general one focuses on so-called close
to diagonal approximants where $N_{k}, k<K,$ differ from $N_K$ by a small amount and one may typically 
use the restriction $|N_K-N_k| \leq 1$ or 2.

If the critical point $x_c$ is known exactly (or very accurately) one may try to obtain improved numerical
estimates for the exponents by forcing the differential equation (\ref{eq:DA})  to have a singular point at $x_c$, that is 
one may look at {\em biased differential approximants}.   We have developed a new method in which we form 
biased approximants by multiplying the derivatives in (\ref{eq:DA}) by appropriate ``biasing polynomials''. 
This allows us to bias in such a manner that the singularity at $x_c$ is of order $q\leq K$. Let

\begin{equation}\label{eq:bias}
F_k(x) = \left(x\frac{{\rm d}}{{\rm d}x}\right)^k F(x) \quad \textrm{and} \quad G_k(x) = (1-x/x_c)^{q_k} F_k(x),
\end{equation}
where $q_k=\max(q+k-K,0)$.  With this definition we have that $G_k= (1-x/x_c)^qF_k(x)$, while subsequent
lower order derivatives have ``biasing polynomials'' of degree decreasing in steps of 1 (until 0). This choice
corresponds to assuming that the singularity is regular \cite {Ince56}.
Then we form biased differential approximants (BDA) such that
 \begin{equation} \label{eq:BDA}
P(x)+\sum_{k=0}^K \widehat{Q}_{k}(x)G_k(x)   = O(x^{N+1}).
\end{equation}
One can readily generalise this to include further singularities at $x_i$ with the order of each given by $q_i$
and the singularity at $x_i$ need not be given by a monomial as above but could in general be given by the
root of a polynomial $p(x)$ with $p(x)$ used as (a factor in) the ``biasing polynomial''. 
For biased approximants $[N_K,N_{K-1},\ldots,N_0;L]$ still denotes a case in which
the degree of the polynomial multiplying the $k$'th derivative have degree $N_k$ such that
the degrees of $\widehat{Q}_{k}(x)= N_k-q_k$ and the number of unknown coefficients  is now
$$N=L+1 +  \sum_{k=0}^K(N_k-q_k+1).$$
 
\section{Analysis and Results}

The critical point for square lattice self-avoiding walks is not know exactly but  has been estimated to a high level
of accuracy. Series analysis of the self-avoiding polygon series \cite{Clisby12} yielded the
estimate $x_c=0.379052277752(3)$ which recently has been improved by 
Guttmann, Jacobsen and Scullard  \cite{GuttmannMail} to $x_c=0.3790522777552(3)$ and
henceforth we shall adopt the value $x_c=0.3790522777552$.
As mentioned in the introduction numerical estimates for the SAW critical exponent $\gamma$ 
from un-biased DAs are close to the exact value but not quite on target. 
This means that  square lattice  SAW should be the perfect model on which to test
the efficacy of our new approach to biased differential approximants.
The number of square lattice SAW has been calculated up to length $n=79$ and the
generating functions for the metric properties of SAW up to length $n=71$ \cite{Jensen13}.

\subsection{The SAW generating function}

\begin{figure}[ht]
\begin{center}
\includegraphics[scale=0.8]{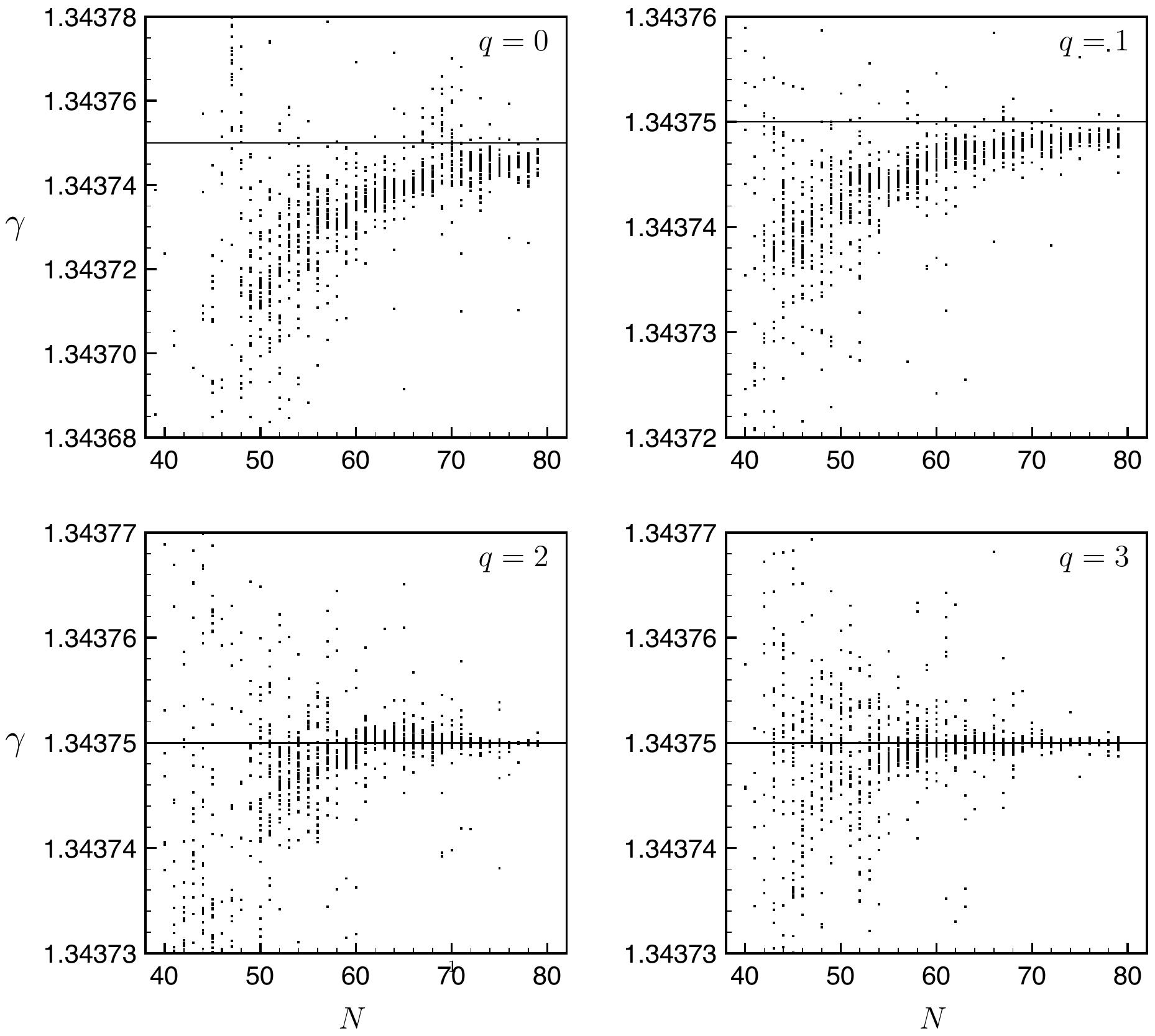}
\end{center}
\caption{\label{fig:gamma}
Estimates of the critical exponent $\gamma$ of square lattice SAW as obtained from 3rd order differential approximants.
The exponent estimates are  plotted versus the
number of terms $N$ used to form the DA. The panels show the results from unbiased DAs ($q=0$) and
DAs biased to have a singularity at $x_c= 0.3790522777552$ with the order of the root changing from $q=1$ to 3.
}
\end{figure}

First we take a look at results obtained from 3rd order differential approximants both un-biased ($q=0$)
and biased with the order of the root at $x_c$ being $q=1,2$ and 3. 
We generate lots of  homogenous  and inhomogenous   $(L=2,4,6,8,10)$ approximants by varying
$N_3$ from 8 to 20 and for each value of $N_3$ we generate all approximants with $|N_3-N_k| \leq 1, k=0,1,2$
(this amounts to more than 1600 approximants for each case). The smallest degree approximants thus utilise
around 40 terms and the highest degree all 79 terms in the SAW generating function.
In \fref{fig:gamma} we plot the resulting estimates of $\gamma$ as s function of $N$.
The panel for $q=0$ (un-biased case) vividly illustrates the frustration the author has had in
analysing this series. As can be seen the estimates exhibit a slow systematic drift  as $N$ is increasing
and ever so slowly they creep towards the exact value (though one might question whether they would
ever get there). Now for $q=1$ the estimates get much closer to the exact value (note the different scale
on the $y$-axis), but again the approach is slow and not quite there yet. The explanation for this
is fairly simple. If one looks at some individual approximants and calculate the roots of $ \widehat{Q}_{K}(x)$
one finds a root that is very close to $x_c$ indicating that the series fells the presence of a confluent 
singularity at $x_c$ and that $q=1$ is therefore not sufficient biasing to pin down the critical behaviour.
So we look to the $q=2$ and 3 cases and now we find a very good convergence of the estimates to
the exact value. For large value of $N$ there is no discernible difference between the two cases
but for low values of $N$ the $q=3$ biased approximants are superior. We also note that
$ \widehat{Q}_{K}(x)$ does not have a root close to $x_c$ when $q\geq 2$ thus confirming
that the biasing has successfully accounted for the critical behaviour.

\begin{table}[tb]
\caption{\label{tab:gamma} Estimates for the critical exponent $\gamma$ obtained from
$K$th order biased differential approximants. We vary the order of the root at $x_c$
from $q=1$ to $K$ and that at $-x_c$ from $r=0$ to $\min(q,2)$.
}
\vspace{2mm}
\begin{tabular}{rlrlrl} \hline \hline
$[K,q,r]$ & $\gamma$    & $[K,q,r]$ &  $\gamma$    & $[K,q,r]$ &  $\gamma$  \\
\hline
$[2,1,0]$ & $1.34374784(48)$ & $[2,1,1]$ & $1.34374779(50)$ \\
$[3,1,0]$  &  $1.34374830(25)$ & $[3,1,1]$ & $1.34374831(25)$ \\
$[4,1,0]$  &  $1.34374838(39)$ & $[4,1,1]$ & $1.34374832(39)$ \\
$[2,2,0]$ & $1.34375009(15)$ & $[2,2,1]$ & $1.34375008(13)$ & $[2,2,2]$ & $1.34375008(13)$ \\
$[3,2,0]$  &  $1.34375003(10)$ & $[3,2,1]$ & $1.34375004(12)$  & $[3,2,2]$ & $1.34375002(10)$ \\
$[4,2,0]$  &  $1.34374997(15)$ & $[4,2,1]$ & $1.34375000(15)$ & $[4,2,2]$  &  $1.34375000(14)$   \\
$[3,3,0]$  &  $1.343750024(80)$\,\,\, & $[3,3,1]$ & $1.343750025(72)$\,\,\, & $[3,3,2]$  &  $1.343750041(62)$   \\
$[4,3,0]$  &  $1.34375003(13)$ & $[4,3,1]$ & $1.34375004(13)$ & $[4,3,2]$  &   $1.34375004(14)$   \\
$[4,4,0]$  &  $1.343750059(62)$ & $[4,4,1]$ & $1.343750057(84)$ & $[4,4,2]$  &  $1.343750071(92)$   \\  \hline \hline
\end{tabular}
\end{table}

As noted in the introduction the SAW generating function also has a singularity on the
negative real axis at $x=x_-=-x_c$  with critical exponent of $1/2$.  We can easily
bias the approximants to include this singularity as well. In forming the biased 
appproximants (\ref{eq:BDA}) we simply modify the biasing done in (\ref{eq:bias}) 
to include the terms $(1+x/x_c)^{r_k}$ in $G_k(x)$, where $r_k=\max(r+k-K,0)$.
This corresponds to biasing with a root of order $r$  at $-x_c$. One can
naturally vary $q$ and $r$ independently. As per above we calculate a large set 
of biased approximants in the cases of $K=2,3,$ and 4 and $q=1$ to $K$.
To obtain estimates for $\gamma$ we extract the relevant critical exponent   
(recall there are $q$ critical exponents at $x_c$ in a biased approximant)
 and we only results from approximants with $N\geq70$. The exponents are sorted, 
 next we remove or ``clip'' the bottom and top 10\% of this data\footnote{This procedure
 will automatically eliminate any spurious outliers some of which will always be present
 when so many approximants are generated.}
 and calculate the mean and standard deviation
of the remaining exponents. The results are displayed in \tref{tab:gamma}
where we list the estimates obtained from the mean. In parenthesis we 
show the standard deviation as an `error' estimate on the last 
two digit.  So the result $1.34375008(13)$ says that the mean was $1.34375008$
and the standard deviation was $0.00000013$. Note that we are {\em not} claiming 
that the standard deviation is the true error estimate. To be on the safe side one should 
use an error bound of at least two to three times the  error estimate and 
one should always check as done here by plotting exponent estimates versus $N$
that there is no systematic drift in the exponent estimates. The results in \tref{tab:gamma}
are in complete accordance with the observation from above that for $q\geq 2$
the biased estimates are spot-on the exact value.  We are in fact confident
in saying that the value of $\gamma$ can be confirmed to at least significant 8 digits. 
We note that there in no improvement to the estimates from $r=0$ to $r=1$ or 2,
so biasing the approximants at $-x_c$ makes no difference at all to the estimates for $\gamma$.
Finally, it is worth mentioning  that changing the `clipping' to say 5\%  or 20\% has next to no
effect of the mean but does change the standard deviation somewhat. In conclusion
we claim that our new biased differential approximant method yields a conservative estimate 
$\gamma=1.3437500(3)$.

\begin{table}
\caption{\label{tab:negxc}  Estimates for the critical exponent $\gamma_-$ at $x=-x_c$ obtained from
3rd order biased differential approximants. We vary the order of the root at $x_c$
from $q=0$ to 3 and that at $-x_c$ from $r=1$ to $3$.
}
\vspace{2mm}

\begin{tabular}{rlrlrl} \hline \hline
$[K,q,r]$ & $\gamma_-$    & $[K,q,r]$ &  $\gamma_-$    & $[K,q,r]$ &  $\gamma_-$   \\
\hline
$[3,0,1]$  &  $0.5000014(13)\;\;$ & $[3,0,2]$ & $0.5000011(21)\;\;$ & $[3,0,3]$ & $0.5000011(20)$ \\
$[3,1,1]$  &  $0.5000014(12)$ & $[3,1,2]$ & $0.5000009(29)$ & $[3,1,3]$ & $0.5000014(24)$ \\
$[3,2,1]$  &  $0.5000013(10)$ & $[3,2,2]$ & $0.5000008(20)$ & $[3,2,3]$ & $0.5000008(26)$ \\
$[3,3,1]$  &  $0.5000012(12)$ & $[3,3,2]$ & $0.5000015(27)$ & $[3,3,3]$ & $0.5000010(23)$ \\
  \hline \hline
\end{tabular}
\end{table}

Next we perform a similar analysis but for the exponent $\gamma_-$ at the singularity $x_-=-x_c$.
The results for $K=3$ are listed in \tref{tab:negxc}. As can be seen the exponent value
$\gamma_-=1/2$ can be confirmed to at least 5 significant digits and there is no change in
the estimates as one changes $r$ and $q$ (even the $q=0$ case gives similar estimates). 
We note that an un-biased analysis gives $\gamma_- =0.500015(15)$, that is, an
order of magnitude less accurate.

\begin{figure}[h]
\begin{center}
\includegraphics[scale=0.82]{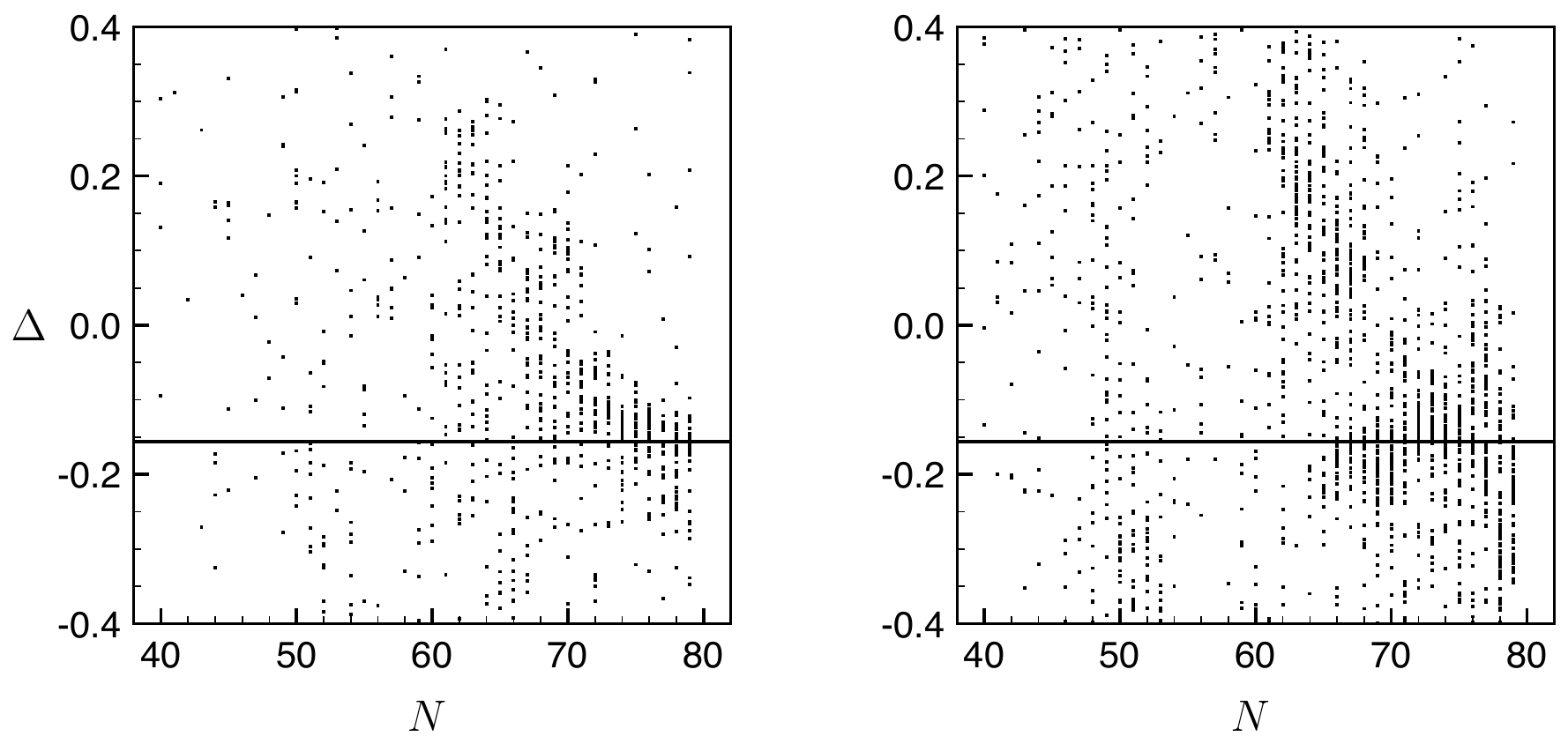}
\end{center}
\caption{\label{fig:subdom}
Estimates of the  sub-dominant critical exponent for square lattice SAW. The left panel
is for $K=3, q=2$ and the right panel $K=4, q=2$. The straight line corresponds to 
the value $\Delta =3/2$ for the leading non-analytic correction to scaling exponent.
}
\end{figure}

Finally, we address the question of estimating the sub-dominant exponents $\gamma_s$. 
The leading correction-to-scaling exponent is expected to have the
exact value $\Delta = 3/2$  \cite{Nienhuis82,Nienhuis84}. In \cite{Caracciolo05} the
correction-to-scaling exponents for SAW was estimated numerically
and a detailed and careful analysis  unequivocally confirmed that   $\Delta = 3/2$ is correct.
But can we confirm this value from a biased differential approximant analysis? For
$q\geq 2$ there are $q$ exponent values at $x_c$ and one might hope that 
of these (just the second exponent in the case $q=2$) will equal $\gamma_s=-\gamma+\Delta=5/32$.
\Fref{fig:subdom} shows the estimates for $-\gamma_s$ plotted versus $N$
in the cases of $q=2$ and $K=3$ and 4.
The straight line is the value $\gamma-\Delta=-5/32$ and as can be seen the
estimates are certainly consistent with this value for large enough $N$. Unfortunately
that is all one can say. So biasing does not lead to an accurate estimate for $\Delta$.
This does not change for higher values of $q$ where similar plots can be made (the
third and fourth exponents are of no use whatsoever).

\subsection{Metric properties}

Finally we analyse the generating functions for three
metric properties. Firstly $\RGf_e (x)$ the mean-square end-to-end distance  of an $n$ step SAW,
secondly $\RGf_m (x)$ the mean-square distance of a monomer (or vertex) from the end-points and thirdly
$\RGf_g (x)$ the mean-square radius of gyration of the monomers of the SAW. Precise definitions
 of the relevant quantities can be found in say \cite{Caracciolo05}. Suffice to that these
 functions are expected to have the critical behaviour
 
 \begin{eqnarray}
 \RGf_e (x) &=& \sum_{n} c_n \ave{R^2_e}_n x^n  \qquad \qquad \propto \;\;   (1-x/x_c)^{-(\gamma+2\nu)}, \\
\RGf_m (x) &=& \sum_{n} (n+1)c_n \ave{R^2_m}_n x^n \quad \propto \;\;   (1-x/x_c)^{-(\gamma+2\nu+1)}, \\
   \RGf_g (x) &=& \sum_{n} (n+1)^2 c_n \ave{R^2_g}_n x^n  \quad  \propto \;\;    (1-x/x_c)^{-(\gamma+2\nu+2)}, 
 \end{eqnarray}
where the factors under the sum ensure that the coefficients are integer 
valued. Here $\nu$ is another critical exponent with value $\nu=3/4$   \cite{Nienhuis82,Nienhuis84}.

\begin{figure}[t]
\begin{center}
\includegraphics[scale=0.72]{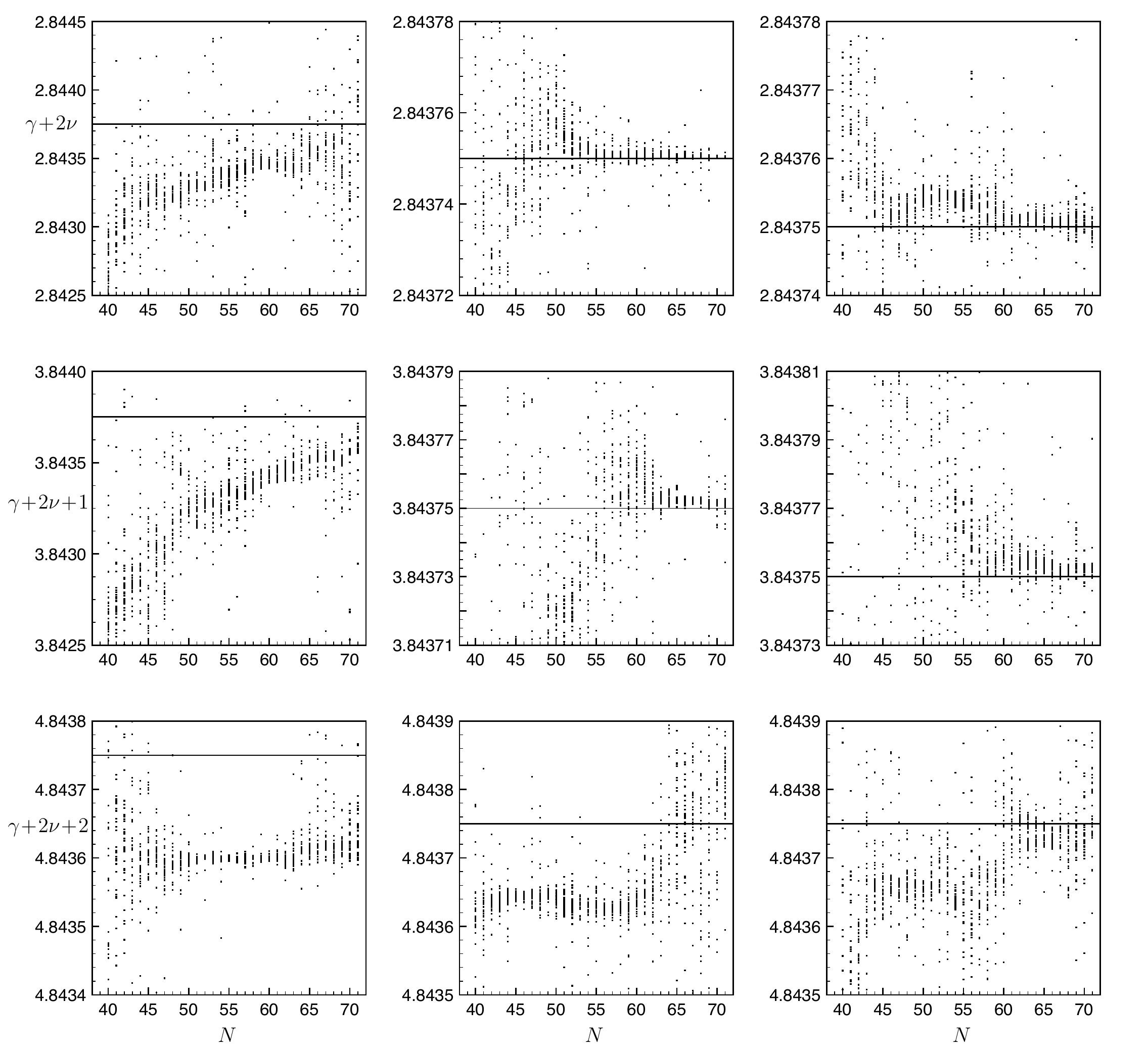}
\end{center}
\caption{\label{fig:metric}
Estimates of the  critical exponents for the 
metric properties of square lattice SAW. In each case the estimates are obtained from 3rd order
differential approximants. From left to right are unbiased estimates
and then biased estimates with $q=2$ and 3. From top to bottom are estimates
to the generating function for the mean-square end-to-end distance, the mean-square monomer-to-end distance
and the mean-square radius of gyration. 
}
\end{figure}

\Fref{fig:metric} displays results obtained from 3rd order differential approximants both un-biased ($q\!=\!0$)
and biased with the order of the root at $x_c$ being $q\!=\!2$ and 3, respectively. 
The left-most panels which are for the un-biased case again vividly illustrates  that traditional differential approximants
cannot quite reproduce the exact values of the critical exponents. 
As  was the case for the SAW generating function the estimates from the mean-square end-to-end distance $\RGf_e (x)$  (top panel)
and the mean-square monomer-to-end distance $\RGf_m (x)$  (middle panel) exhibit a slow systematic drift  as $N$ is increasing
and ever so slowly they approach the exact value. The estimates from the mean-square radius-of-gyration
$\RGf_g (x)$  (bottom panel) are particularly `poor' and seem to settle at a value somewhat below the
exact value.  When we look at the $q\!=\!2$ and 3 biased approximants   we find a very good 
convergence of the estimates from $\RGf_e (x) $ and $\RGf_m (x) $ to
the exact exponent values (note in particular the finer scale along the $y$-axis). For $\RGf_g (x)$ the
estimates for large $N$ are now clearly consistent with the exact value but the scatter among the
exponent estimates is quite pronounced and much larger than for the other two series.

 \begin{table}[htb]
\caption{\label{tab:metric} Estimates for the critical exponents of the generating functions
for the mean-square  end-to-end distance, mean-square  monomer-to-end distance and mean square radius of gyration as obtained from
$K$th order biased differential approximants.  
}
\vspace{2mm}
\begin{center}
\begin{tabular}{rlrlrl} 
\hline \hline
\multicolumn{6}{c}{Mean-square end-to-end distance} \\ \hline
$[K,q,r]$ & $\gamma\!+\!2\nu$    & $[K,q,r]$ &  $\gamma\!+\!2\nu$    & $[K,q,r]$ &  $\gamma\!+\!2\nu$    \\
\hline
$[2,1,0]$ & $2.843679(34)$ & $[2,2,0]$ & $2.84375113(29)$ \\
$[3,1,0]$  &  $2.843683(14)$ & $[3,2,0]$ & $2.84375050(14)$ & $[3,3,0]$ & $2.84375081(61)$  \\
$[4,1,0]$  &  $2.8436984(58)$ & $[4,2,0]$ & $2.84375033(29)$ &  $[4,3,0]$ & $2.84375064(24)$ \\ \hline \hline \\
\hline \hline
\multicolumn{6}{c}{Mean-square monomer-to-end distance} \\ \hline
$[K,q,r]$ & $\gamma\!+\!2\nu\!+\!1$    & $[K,q,r]$ &  $\gamma\!+\!2\nu\!+\!1$    & $[K,q,r]$ &  $\gamma\!+\!2\nu\!+\!1$    \\
\hline
$[2,1,0]$ & $3.843649(31)$ & $[2,2,0]$ & $3.84375268(60)$ \\
$[3,1,0]$  &  $3.8436775(78)$ & $[3,2,0]$ & $ 3.84375198(47)$ & $[3,3,0]$ & $3.8437524(13)$  \\
$[4,1,0]$  &  $3.8436803(72)$ & $[4,2,0]$ & $3.8437525(31)$ & $[4,3,0]$ & $3.8437524(14)$ \\  \hline \hline \\
\hline \hline
\multicolumn{6}{c}{Mean-square radius of gyration} \\ \hline 
$[K,q,r]$ & $\gamma\!+\!2\nu\!+\!2$    & $[K,q,r]$ &  $\gamma\!+\!2\nu\!+\!2$    & $[K,q,r]$ &  $\gamma\!+\!2\nu\!+\!2$    \\
\hline
$[2,1,0]$ & $4.8437055(82)$ & $[2,2,0]$ & $4.843691(64)$ \\
$[3,1,0]$  &  $4.8437183(92)$ & $[3,2,0]$ & $4.84375(10)$ & $[3,3,0]$ & $4.843734(24)$  \\
$[4,1,0]$  &  $4.843714(17)$ & $[4,2,0]$ & $4.843673(76)$  & $[4,3,0]$ & $4.843743(66)$ \\
  \hline \hline
\end{tabular}
\end{center}
\end{table}

In \tref{tab:metric}  we list estimates for the critical exponents of the three
metric generating functions as  obtained from the averaging procedure described
above except in this case we use the approximants with $N\geq 63$. The data
naturally confirm the qualitative observations made from  \fref{fig:metric} but
now in a quantitative manner. The estimates form un-biased approximants (column 2)
are systematically lower than the exact value with `error-bounds' that does not
quite include the exact value. For the $q=2$ and 3 cases the estimates 
for $\RGf_e (x)$ our estimate  $\nu=0.7500002(4)$ gives agreement with the exact exponent 
value $\nu=3/4$ to 7 digits and thus provide a high accuracy confirmation of the exact value. 
The estimates for  $\RGf_m (x)$ are generally an order of magnitude less accurate but again
confirm the exact values. Finally, for  $\RGf_g (x)$ we obtain estimates
fully consistent with the exact value but now even less accurate.
Curiously, the estimates for the un-biased case appear to be more accurate
(as one can also confirm from \fref{fig:metric}), but this is quite clearly 
a `false' convergence.

Finally we calculated biased estimates for the exponents $\gamma_-$ at $-x_c$. Here
we just quote results for $K=3$ and 4 and $q=2, r=1$:

\begin{center}
\begin{tabular}{lll}
$\; \RGf_e (x)\;\;\;$  & $K=3\!: \;\; \gamma_-=\;0.500027(63)\;\;\;\;$ & $K=4\!: \;\; \gamma_- =\; 0.500050(42)$ \\ \\
$\RGf_m (x)$ & $K=3\!: \;\;  \gamma_- =\;-1.99952(11) $ & $K=4\!: \;\;  \gamma_- =\;-1.99952(16)$ \\ \\
$\RGf_g (x)$ & $K=3\!: \;\;  \gamma_- =\;-2.99965(13) $ & $K=4\!: \;\;  \gamma_- =\;-2.99968(40)$ 
 \end{tabular}
\end{center}
These estimates are clearly consistent with the exact values $\gamma_- = 1/2$ for $\RGf_e (x)$,
$\gamma_- = -2$ for $\RGf_m (x)$ and $\gamma_- = -3$ for $\RGf_g (x)$.

\section{Summary and Outlook}

We  made use of the fact that the critical point for square lattice self-avoiding walks
is known very accurately to perform a numerical analysis of the SAW series using a new
method for biased differential approximants.  Our new method is a major advance 
since it permits us to bias a  differential approximant such that is has a singularity
of order $q$ (up to the order of the underlying differential equation).
From the analysis with $q\geq 2$ we obtained exponent
estimates in total agreement with the conjectured exact values $\gamma=43/32$ and $\nu=3/4$.
In the case of the SAW generating function our conservative estimate  $\gamma=1.3437500(3)$ 
confirmed the exact value to at least 8 significant digits. This estimate is several orders of magnitude 
more accurate than the estimate from un-biased  differential approximants
and eliminates a long-standing minor discrepancy between exact and numerical results.
In the case of the metric properties we obtain similar impressive estimates from the analysis
of the mean-square end-to-end distance $\RGf_e (x)$ where our estimate $\nu=0.7500002(4)$ 
confirmed the value of $\nu$ to 7 significant digits. The results for
$\RGf_m (x)$ was not quite so impressive being an order of magnitude less accurate.
The only slightly peculiar case was $\RGf_g (x)$ where the un-biased estimates appears
to to be quite accurate but wrongly gives an exponent estimate below the exact value while 
the biased estimates are consistent with the exact value but much less accurate. So in
this case the biased exponents at least gives the correct answer which obviously is better
than an accurate but wrong estimate. We also
estimated the critical exponents $\gamma_-$  at $-x_c$
and found exponent consistent with the  exact values $\gamma_- = 1/2$ for SAW and $\RGf_e (x)$,
$\gamma_- = -2$ for $\RGf_m (x)$ and $\gamma_- = -3$ for $\RGf_g (x)$. Finally, we
tried to estimate the leading non-analytic correction-to-scaling exponent $\Delta$ from the
biased exponent results when $q\geq 2$. We found results consistent with the 
exact conjectured value $\Delta=3/2$ but somewhat disappointingly the estimates were not at all accurate.

The new method of  biased differential approximants that we have introduced is this paper
promises to be very useful in obtaining accurate exponent estimates in cases where the
critical point(s) of a power-series is known exactly or very accurately. An obvious application
is to various percolation series which have been notoriously difficult to analyse.

An interesting question is whether the method can be used to `reverse engineer' more
accurate estimates for critical points. In many cases (particularly 2D systems) critical
exponents are known exactly but not so the critical points. Would it be possible to obtain
improved estimates for the critical point say by calculating biased estimates for critical exponents
while varying the biasing value of $x_c$. One might expect (or at least hope) that the 
biasing value of $x_c$ yielding an exponent estimate equal to the exact value should be
very close to the true critical point. A preliminary trial run using the square lattice SAP generating 
function confirmed the overall picture but did not give an improved estimate for $x_c$.
But that is hardly surprising given that the un-biased estimates for $x_c$ and $\alpha$
are already extremely accurate.

\label{sec:summary}
   \section*{Acknowledgements}
This work was supported by an award under the Merit Allocation Scheme on the NCI National Facility  
 and  by  funding under the Australian Research Council's Discovery Projects scheme by the grant
DP140101110.

 \section*{References}


\end{document}